\documentclass{pprai}

\usepackage[utf8]{inputenc}
\usepackage[T1]{fontenc}
\usepackage{graphicx}
\usepackage{amsmath}
\usepackage{amsthm}
\usepackage{txfonts}
\usepackage{url}
\usepackage{algpseudocode}
\usepackage{float}
\usepackage{wrapfig}
\usepackage[export]{adjustbox}
\usepackage{hyperref}

\title{Generative Diffusion Models for Fast Simulations of Particle Collisions at CERN}
\headtitle{ Generative Diffusion Models for Fast Simulations of Particle Collisions at CERN}

\author{Mikołaj Kita$^{1}$, Jan Dubiński$^{1, 2}$, Przemysław Rokita$^{1}$, Kamil Deja$^{1, 2}$}
\headauthor{M.Kita, J.Dubiński, P.Rokita, K.Deja}
\affiliation{%
  $^1$Warsaw University of Technology\\
  $^2$IDEAS NCBR\\
  mikolajkita@gmail.com}
\keywords{diffusion models, particle simulations, ALICE, CERN}

\begin{document}
\maketitle

\begin{abstract}
In High Energy Physics simulations play a crucial role in unraveling the complexities of particle collision experiments within CERN's Large Hadron Collider. Machine learning simulation methods have garnered attention as promising alternatives to traditional approaches. While existing methods mainly employ Variational Autoencoders (VAEs) or Generative Adversarial Networks (GANs), recent advancements highlight the efficacy of diffusion models as state-of-the-art generative machine learning methods. We present the first simulation for Zero Degree Calorimeter (ZDC) at the ALICE experiment based on diffusion models, achieving the highest fidelity compared to existing baselines. We perform an analysis of trade-offs between generation times and the simulation quality. The results indicate a significant potential of latent diffusion model due to its
rapid generation time.
\end{abstract}

\section{Introduction}
The European Organization for Nuclear Research (CERN) is one of the world's most renowned centers for scientific research in High-Energy Physics (HEP), where the Large Hadron Collider (LHC)~\cite{Evans:2008zzb}, the world's largest and most powerful particle accelerator is located. To understand what happens during these collisions, sophisticated simulations of the detectors inside the LHC are generated. The majority of these simulations are founded on Monte Carlo methods~\cite{Geant4}, which yield results of high quality, albeit with a significant computational expense. In response, researchers investigate  strategies to optimize resource use, including the application of generative machine learning methods to lessen CPU consumption \cite{CaloGAN}, \cite{e2eSinkhorn}.

Current efforts to develop generative simulations for ZDC include the usage of Variational Autoencoders (VAEs) \cite{Variational_AE} and Generative Adversarial Networks (GANs) \cite{GoodfellowGAN} for simulating particle responses \cite{CaloGAN,CERN_Dataset,e2eSinkhorn}. While they present an improvement over the existing methods based on Monte Carlo, several challenges have been identified, notably the production of indistinct samples in VAEs and an unsatisfactory variety of samples generated by GANs. Methods to enhance the diversity of GAN-generated samples and to diminish the blurriness of VAE-generated images exist \cite{e2eSinkhorn, CERN_Dataset}. Nevertheless, a promising alternative to such approaches is the usage of the newly proposed diffusion models. Originally, diffusion models were employed in image generation \cite{DDPM2020} and have since branched out to other areas like video \cite{video_diffusion}, text-to-audio \cite{audioldm}, and even text generation \cite{gong2023diffuseq}. The demonstrated versatility of diffusion models presents potential for their application in various fields \cite{cholect, application1, application2}, including physics simulations. Therefore, in this work, we introduce a new simulation method for ZDC that leverages diffusion models. Our method achieves better performance on Wasserstein distance evaluation metric for conditional diffusion model and presents significant potential of latent diffusion model due to its rapid generation time.
\section{Related work}
Numerous efforts have been directed towards supplanting the resource-intensive Monte Carlo methods traditionally used for High Energy Physics simulations with more efficient alternatives like generative modeling. 
Most of the proposed solutions are based on GANs, such as \cite{CaloGAN, CERN_Dataset, dubiński2023selectively}, but other methods such as generative autoencoders \cite{e2eSinkhorn} or score-based generative models are also used \cite{Score_gen_models}. The proliferation of those approaches lead to the creation of Fast Calorimeter Simulation Challenge \cite{Fast_Challenge} with three high-quality datasets with calorimeter responses to particle collisions.

Apart from calorimeters simulations, diffusion models find extensive application in physics simulations. Their utilization spans from the reconstruction of flow fields \cite{FlowField} to the generation of jets within the realm of HEP \cite{PC-JeDi,FPCD}.

\section{Zero Degree Calorimeter simulation}
The neutron Zero Degree Calorimeter (ZDC) is a quartz-fiber spaghetti calorimeter, utilizing the principle of detecting Cherenkov light produced by charged particles of the shower in silica optical fibers \cite{NeutronZDC}. Every alternate fiber is directed towards a photomultiplier (PMTc), with the rest of the fibers being grouped into bundles that lead to four distinct photomultipliers (PMT1 to PMT4). The design allows for precise measurement of neutron energy in heavy ion collisions at the CERN LHC.

\begin{figure}[H]
	\centering \includegraphics[width=0.5\linewidth]{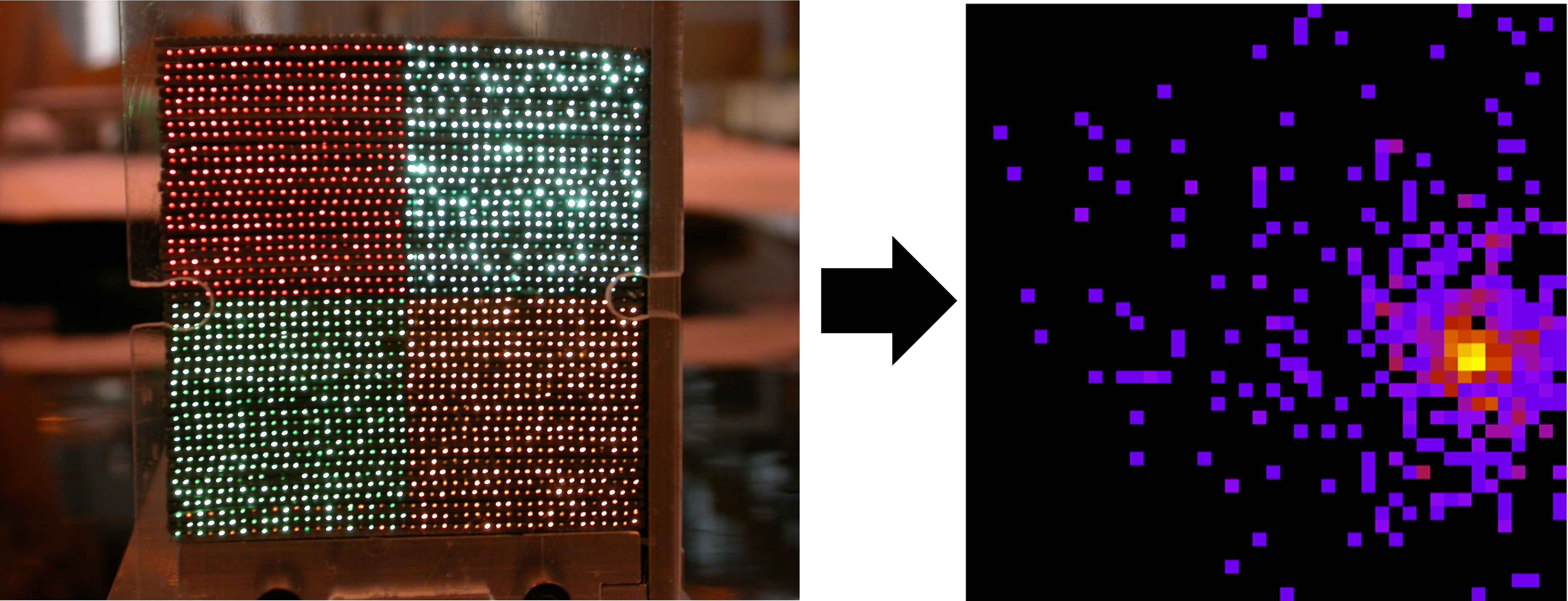}
	\caption{ZDC cross-section with visible optic fibres grid (left) represented  as a one-channel image (right)}
    \label{fig:ZDC-Conversion}
\end{figure}

The response of the ZDC is treated as a one-channel image composed of a 44 by 44 pixel matrix, wherein the value of each pixel represents the count of photons deposited in an individual fibre, which we visualize in Fig.~\ref{fig:ZDC-Conversion}. 

\vspace{-10pt}

Every response is produced by a  particle, which is characterized by attributes: mass, energy, charge, momenta, and primary vertex. The correlation between particle attributes and responses is non-deterministic. Consequently, a single set of nine attributes may yield a variety of different responses, necessitating evaluations across a distribution of generated images. The dataset consists of 300000 image-particle pairs from GEANT4 simulation tool.

\section{Method}
\begin{figure}[H]
    \centering \includegraphics[width=0.7\linewidth]{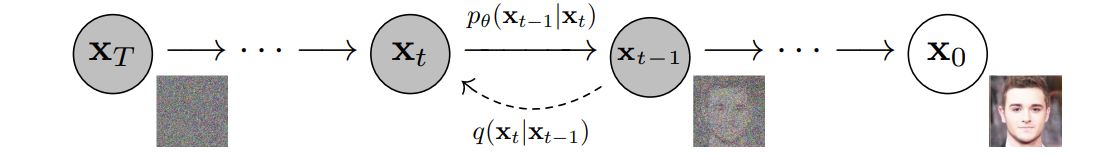}
    \caption{\label{fig:forward_reverse_diffusion_pass}Graphical presentation of forward and reverse diffusion pass. Image from~\cite{DDPM2020}}
\end{figure}

Diffusion models consist of \textit{forward diffusion pass} $q(x_{1:T}|x_{0})$, which is described by a Markov chain that progressively adds noise to the original data, and a \textit{reverse diffusion pass} $p_{\theta}(x_{0:T})$, which aims to reconstruct the original image from the noised data with a trainable model, as referenced in Fig.~\ref{fig:forward_reverse_diffusion_pass}.

\begin{wrapfigure}{r}{0.45\linewidth}
    \centering \includegraphics[width=0.99\linewidth]{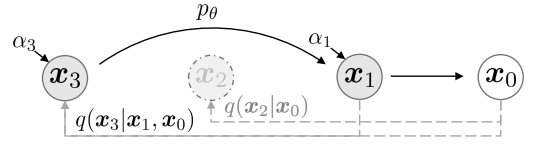}
    \caption{\label{fig:DDIM_process} Accelerated generation with DDIM sampling method. Image from~\cite{DDIM2020}}
\vspace{-8pt}
\end{wrapfigure}

An alternative and faster method to sample images from diffusion models is proposed in \cite{DDIM2020}. The authors reinterpret the reverse diffusion process as a non-Markovian process, which allows for omission of intermediate steps in the denoising proccess, enabling faster generation of samples as presented in Fig.~\ref{fig:DDIM_process}

We use a diffusion model based on UNet architecture \cite{2015unet}. To introduce conditioning on the particle parameters, we adapt the text-conditioning of the text-to-image diffusion models. Particle characteristics $y$ are multiplied by a linear layer $\tau_\theta(y)$ to function as an embedding for physical conditions. This allows the model to focus on specific features of the generated data based on the particle data input, enhancing contextual relationships between them:
\begin{equation}
\label{cross_attention}
\text{Attention}(Q, K, V) = \text{softmax}\left(\frac{QK^T}{\sqrt{d}}\right) \cdot V,
\end{equation}
where 
\begin{equation}
Q = W^{(i)}_Q \cdot \phi_i(z_t), \quad K = W^{(i)}_K \cdot \tau_\theta(y), \quad V = W^{(i)}_V \cdot \tau_\theta(y). 
\end{equation}
where $W^{(i)}_Q$, $W^{(i)}_K$, and $W^{(i)}_V$ are learnable matrix projections for query, key and value \cite{LDM}.

During additional experiments, we also investigate Latent Diffusion Models~\cite{LDM}, which run the diffusion process in the latent space of an image autoencoder.
\section{Experiments and  Results}

\subsection{Evaluation}
 
The evaluation process unfolds in two phases. Initially, a test dataset comprising 60,000 images from GEANT4 simulation tool is utilized. Five channel values, corresponding to PMT1 to PMT4 and PMTc, are extracted from each image. Subsequently, a diffusion model generates a comparable set of 60,000 images. The same procedure for channel extraction is applied to these images. The critical step involves calculating the disparity between each pair of corresponding channel distributions using the Wasserstein Distance, yielding five distinct distances. The final step in the evaluation is to compute the average of these five distances. The primary goal is to attain the lowest possible average.

\subsection{Results}
Results are based on generating 60000 images for two sampling methods: DDPM \cite{DDPM2020} and DDIM \cite{DDIM2020} with baseline methods from \cite{CERN_Dataset}.
When comparing sampling methods, the DDIM method shows a slight advantage at lower inference steps, while the DDPM method notably excels at the higher end of the spectrum.
\vspace{-10pt}
\begin{table}[H]
\centering
\caption{Conditional Diffusion Model results based on different sampling methods and number of inference steps, evaluated on a single GeForce RTX 3090 GPU.}
\begin{adjustbox}{width=1\textwidth}
\begin{tabular}{lrrrrrrrr}
\hline
\textbf{Method} &\textbf{Inference steps}& \textbf{Mean WS} & \textbf{ch1} & \textbf{ch2} & \textbf{ch3} & \textbf{ch4} & \textbf{ch5} & \textbf{Time (minutes)} \\
\hline
GAN & - & 8.2 & 4.4 & 5.5 & 7.3 & 9.1 & 15.0 & <1 \\
VAE & - & 6.4 & 4.6 & 5.2 & 4.2 & 9.1 & 13.7 & <1 \\
end2end SAE & - & 6.3 & 4.2 & 5.0 & 4.0 & 4.0 & 13.6 & <1 \\
DCGAN + selective div increase & - & 4.5 & 2.2 & 4.0 & 4.4 & 6.2 & 8.0 & <1 \\
\hline
DDIM & 20 & 9.9 & 5.9 & 5.5 & 6.2 & 6.0 & 25.8 & 6 \\
DDIM & 50 & 6.4 & 3.9 & 3.4 & 3.9 & 3.0 & 17.8 & 15 \\
DDIM & 100 & 21.6 & 13.9 & 13.2 & 14.0 & 12.7 & 54.3 & 31 \\
DDPM & 20 & 10.1 & 6.5 & 6.6 & 7.0 & 7.3 & 23.0 & 6 \\
DDPM & 50 & 7.2 & 4.6 & 4.8 & 4.6 & 5.1 & 16.9 & 15 \\
DDPM & 100 & 4.7 & 3.0 & 3.2 & 2.9 & 3.2 & 11.0 & 31 \\
DDPM & 250 & 2.1 & 1.5 & 1.5 & 1.4 & 1.6 & 4.5 & 56 \\
DDPM & 500 & 1.2 & 0.9 & 1.1 & 0.8 & 1.0 & 2.1 & 109 \\
DDPM & 1000 & 1.8 & 1.7 & 1.8 & 1.9 & 1.0 & 2.4 & 215 \\
\hline
\end{tabular}
\end{adjustbox}
\label{tab:conditional_diffusion_model_results}
\end{table}
\vspace{-20pt}

\begin{figure}[H]
    \centering
    \begin{minipage}{0.47\linewidth}
        \centering
        \includegraphics[width=\linewidth]{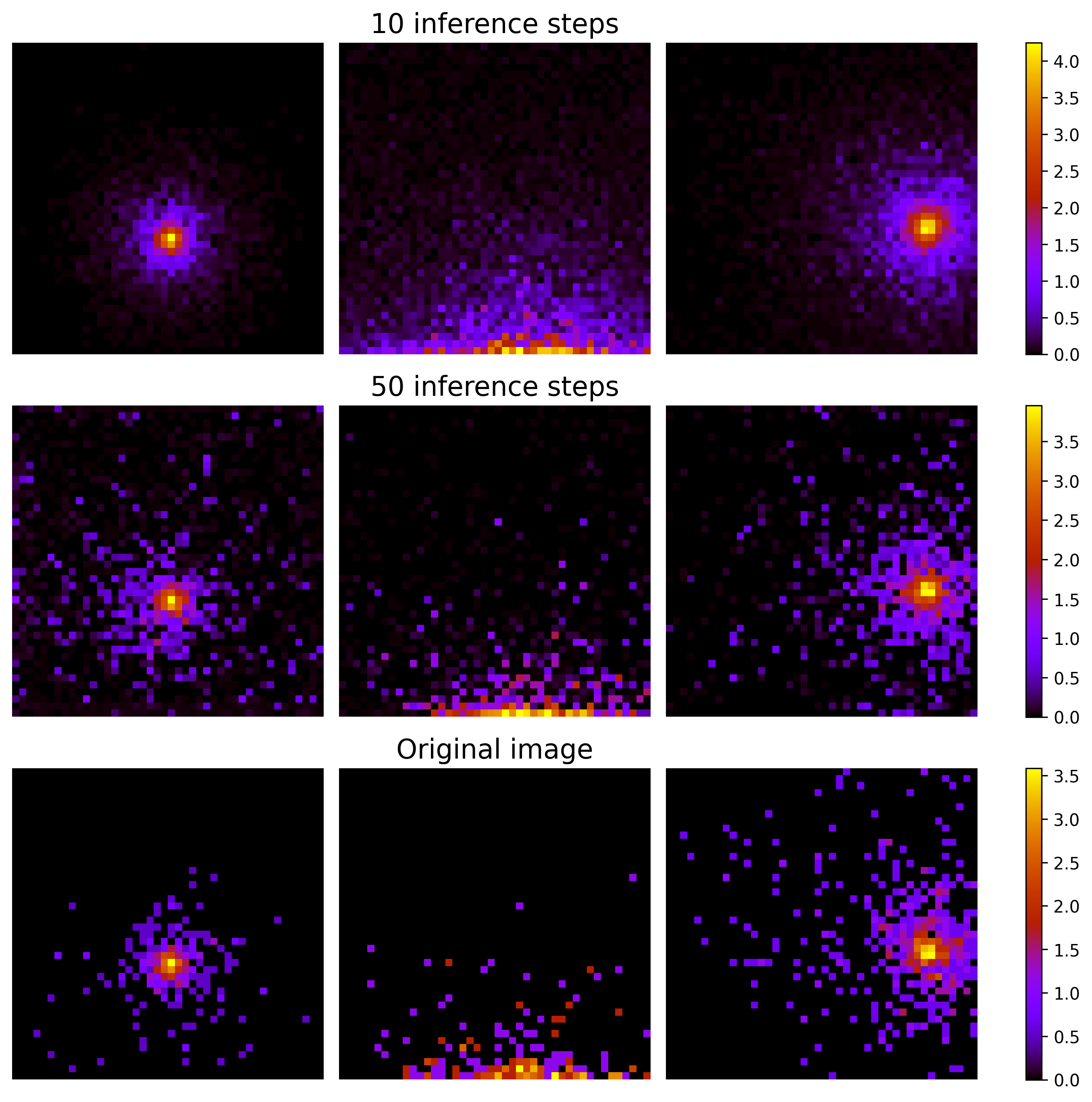}
        \caption{Conditional Diffusion Model generations from the DDIM sampler at different number of inference steps}
        \label{fig:Cond_DDIM}
    \end{minipage}
    \hspace{0.03\linewidth} 
    \begin{minipage}{0.47\linewidth}
        \centering
        \includegraphics[width=\linewidth]{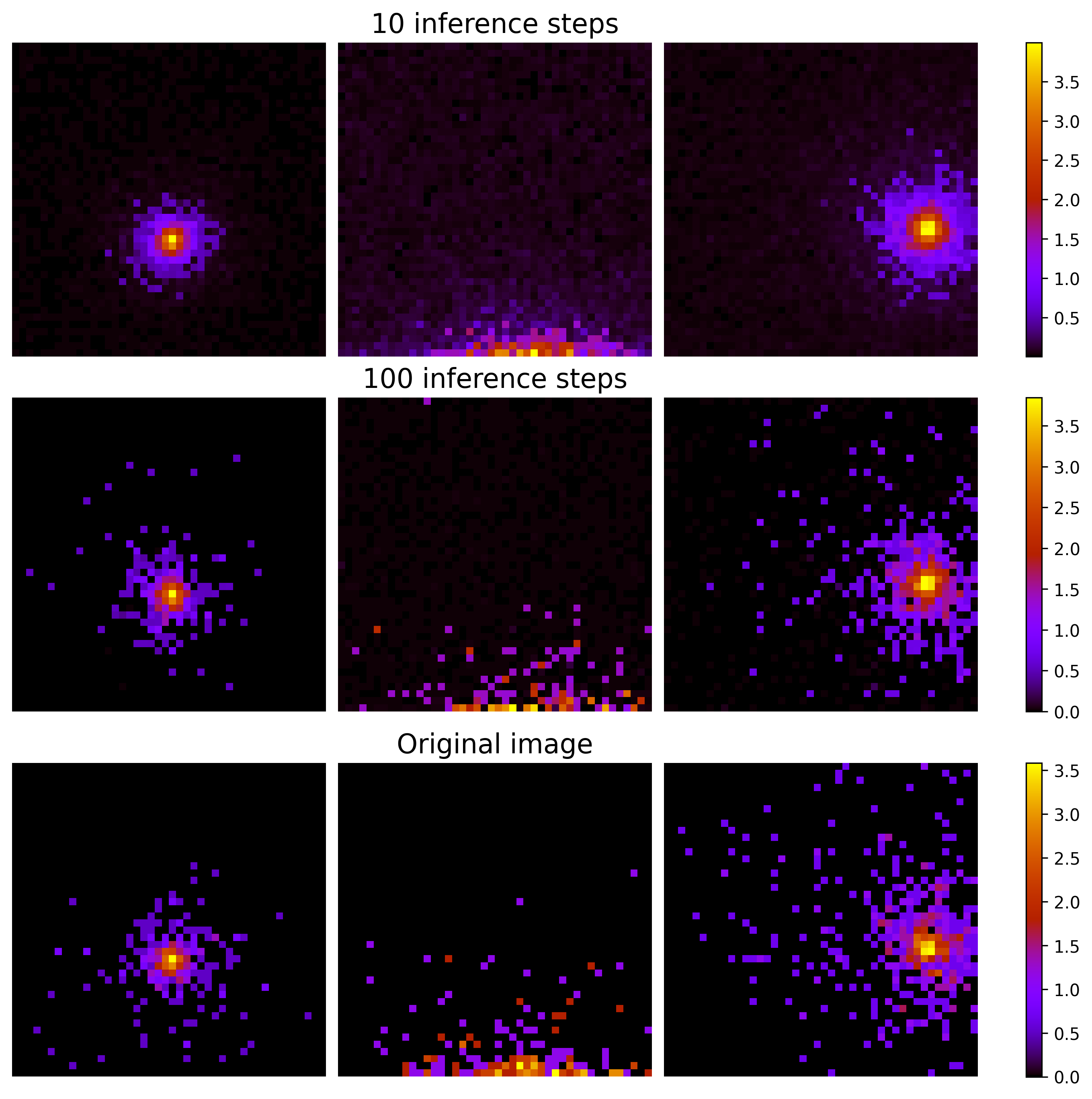}
        \caption{Conditional Diffusion Model generations from the DDPM sampler at different number of inference steps}
        \label{fig:Cond_DDPM}
    \end{minipage}
\end{figure}
The average Wasserstein distance diminishes up to a certain number of inference steps, before it starts to rise. This turnaround happens more swiftly with the DDIM sampler and more gradually with the DDPM sampler.

The progressive decrease in Wasserstein Distance can be seen in Fig.~\ref{fig:Cond_DDIM} and Fig.~\ref{fig:Cond_DDPM}. Images that initially appear slightly blurred at 10 inference steps gradually achieve a quality that is nearly indistinguishable from the original.
\subsection{Additional Experiments}
The superior performance exhibited by conditional diffusion models is accompanied by the substantial cost of increased computational time. In response to this challenge, we have chosen to augment the model with the integration of a Latent Diffusion Model \cite{LDM}. The core concept of LDM involves the integration of an autoencoder that compresses the high-dimensional pixel space into a more computationally manageable latent representation, thus reducing the complexity and resource demands of both training and generative process. We used a conditional VAE with scaling factor 0.18215 and latent space with dimensions [4,5,5].
\begin{table}[H]
\centering
\vspace{-12pt}
\caption{Comparison of diffusion models on Wasserstein Distance and Time performance on a GeForce RTX 3090 GPU}
\begin{adjustbox}{width=1\textwidth}
\begin{tabular}{crrrrrrr}
\hline
\textbf{Model} & \textbf{Average WS} & \textbf{WS ch1} & \textbf{WS ch2} & \textbf{WS ch3} & \textbf{WS ch4} & \textbf{WS ch5} & \textbf{Time (minutes)} \\
\hline
Latent Diffusion Model & 12.6 & 5.8 & 8.8 & 9.1 & 10.5 & 28.6 & <1 \\
Conditional Diffusion Model & 1.2 & 0.9 & 1.1 & 0.8 & 1.0 & 2.1 & 109 \\
\hline
\end{tabular}
\end{adjustbox}
\label{tab:model_performance}
\vspace{-12pt}
\end{table}
As shown in Tab.~\ref{tab:model_performance}, the Latent Diffusion Model presents significant potential due to its rapid generation times. However, it still requires more development work to reach the performance benchmarks set by other models.
\section{Conclusions}
In this paper, we apply generative diffusion models to the problem of simulating the ZDC calorimeter at the ALICE experiment at CERN. We demonstrate the capabilities of conditional diffusion models, outperforming existing state-of-the-art models by achieving lower Wasserstein Distances and generating high-quality, realistic images. Moreover, we show that the Latent Diffusion Model presents significant potential due to its rapid generation times. However, it still requires more development work to reach the performance benchmarks set by other models, which we leave as future work.

\section*{Acknowledgments}
 
This research was funded by National Science Centre, Poland grants: 2020/39/O/ST6/01478 and 2022/45/B/ST6/02817. This research was supported in part by PLGrid Infrastructure grants: PLG/2023/016393, PLG/2023/016361, PLG/2023/016278.

\vspace{-10pt}
{\small
\bibliography{pprai}
\bibliographystyle{pprai}
}

\end{document}